 \documentclass[preprint2]{aastex}

\usepackage{array}
\usepackage{rotating}
\usepackage{graphics}
\newcommand{\hd}{HD\,148937}
\newcommand{\xmm}{{\sc XMM}\emph{-Newton}}
\newcommand{\ro}{{\sc ROSAT}}
\newcommand{\lo}{$\lambda$}
\newcommand{\ha}{H$\alpha$}
\newcommand{\hb}{H$\beta$}
\newcommand{\hg}{H$\gamma$}
\newcommand{\hei}{He\,{\sc i}}
\newcommand{\heii}{He\,{\sc ii}}
\newcommand{\niii}{N\,{\sc{iii}}}
\newcommand{\niv}{N\,{\sc{iv}}}
\newcommand{\nv}{N\,{\sc{v}}}
\newcommand{\siii}{Si\,{\sc{iii}}}
\newcommand{\siv}{Si\,{\sc{iv}}}

\newcommand{\cii}{C\,{\sc{ii}}}
\newcommand{\ciii}{C\,{\sc{iii}}}
\newcommand{\civ}{C\,{\sc{iv}}}
\newcommand{\oii}{O\,{\sc{ii}}}
\newcommand{\oiii}{O\,{\sc{iii}}}
\newcommand{\ov}{O\,{\sc{v}}}
\newcommand{\kms}{km\,s$^{-1}$}

\newcommand{\msol}{M$_{\odot}$}

\shorttitle{Draft}
\shortauthors{Naz\'e et al.}

\begin{document}

\title{\hd: a multiwavelength study of the third Galactic member of the Of?p class\footnote{Based on observations collected with SMARTS and \xmm, an ESA Science Mission with instruments and contributions directly funded by ESA Member States and the USA (NASA).}}

\author{Ya\"el Naz\'e\altaffilmark{2}}
\affil{Institut d'Astrophysique et de G\'eophysique,
Universit\'e de Li\`ege, All\'ee du 6 Ao\^ut 17, B\^at. B5c, B4000 -
Li\`ege, Belgium}
\email{naze@astro.ulg.ac.be} 

\author{Nolan R. Walborn}
\affil{Space Telescope Science Institute\altaffilmark{3}, 3700 San Martin Drive, Baltimore, MD 21218, USA}

\author{Gregor Rauw\altaffilmark{4}}
\affil{Institut d'Astrophysique et de G\'eophysique,
Universit\'e de Li\`ege, All\'ee du 6 Ao\^ut 17, B\^at. B5c, B4000 -
Li\`ege, Belgium}

\author{Fabrice Martins}
\affil{MPE, Postfach 1312, D-85741 Garching bei Muenchen, Germany
- Universit\'e Montpellier II, GRAAL/CNRS, UMR 5024, place Eug\`ene Bataillon, F-34095 Montpellier, France}

\author{A.M.T. Pollock}
\affil{European Space Agency, \xmm\ Science Operations Centre, European Space Astronomy Centre, Apartado 50727, Villafranca del Castillo, 28080 Madrid, Spain}

\and
\author{Howard E. Bond}
\affil{Space Telescope Science Institute\altaffilmark{3}, 3700 San Martin Drive, Baltimore, MD 21218, USA}

\altaffiltext{2}{Postdoctoral Researcher F.R.S./F.N.R.S.}
\altaffiltext{3}{STScI is operated by the Association of Universities for Research in Astronomy, Inc., under NASA contract NAS5-26555.}
\altaffiltext{4}{Research Associate F.R.S./F.N.R.S.}

\begin{abstract}

Three Galactic O-type stars belong to the rare class of Of?p objects: HD\,108, HD\,191612, and \hd. The first two stars show a wealth of phenomena, including magnetic fields and strong X-ray emission, light variability, and dramatic periodic spectral variability. We present here the first detailed optical and X-ray study of the third Galactic Of?p star, \hd. Spectroscopic monitoring has revealed low-level variability in the Balmer and \heii\,\lo\,4686 lines, but constancy at \hei\ and \ciii\,\lo\,4650. The \ha\ line exhibits profile variations at a possible periodicity of $\sim$7d. Model atmosphere fits yield $T_{eff}=41000\pm2000$K, $\log(g)=4.0\pm0.1$, $\dot M_{sph}\lesssim 10^{-7}$\msol\,yr$^{-1}$ and a surabondance of nitrogen by a factor of four. At X-ray wavelengths, \hd\ resembles HD\,108 and HD\,191612 in having a thermal spectrum dominated by a relatively cool component ($kT$=0.2keV), broad lines ($\ge$1700\kms), and an order-of-magnitude overluminosity compared to normal O stars ($\log [L_{\rm X}^{\rm unabs}/L_{\rm BOL}]\sim-6$).

\end{abstract}


\section{Introduction}

The Of?p category was defined by \citet{wal72} to gather peculiar stars displaying strong \ciii\ emission lines around 4650\,\AA, i.e.\ of an intensity comparable to that of the neighbouring \niii\ lines. Three stars of our Galaxy belong to this class: HD\,108, HD\,191612, and \hd. A few others have now been identified in the Magellanic Clouds \citep{hey92,wal00,mas01}.

In recent years, these peculiar objects have attracted quite a lot of attention. In 2001, it was discovered that HD\,108 presents dramatic line-profile variations of its \ciii, \hei\ and Hydrogen lines \citep{naz01}. These variations seemed recurrent with a timescale of approximately 50--60 years \citep[see also][]{naz06}. \citet{wal03} then reported a similar phenomenon for the spectrum of HD\,191612. However, the period was much shorter (only 538d) and appeared correlated with photometric changes \citep{wal04}. A magnetic field was subsequently identified for this star \citep{donati}, with the 538\,d proposed to be linked to the rotation period. In this scenario, HD\,191612 would be an oblique magnetic rotator, a somewhat evolved version of $\theta^1$ Ori C \citep{donati}. The measured period would correspond to the rotational period of the star, which has been slowed down by the magnetic field, and larger emissions would be detected when the magnetically-confined disk is seen face-on while there are reduced or no contamination by emission when the disk is seen edge-on.

However, the analysis of a dedicated \xmm\ observing campaign did not reveal the dominant signature of a magnetically-confined X-ray emitting plasma: apart from a clear overluminosity, both HD\,108 and HD\,191612 display characteristics typical of massive O-type stars \citep[broad lines, low k$T$; see][]{naz04b, naz07}. Optical observations have further shown HD\,191612 to be a binary, though with a different period \citep[1542d, see][]{how07}. A magnetically-confined disk therefore appears insufficient to explain the full behaviour of those Of?p stars; an additional source of relatively soft X-rays is needed but its origin is still unknown (is it somehow related to binarity, like e.g. colliding wind emission, or is it a more exotic phenomenon?) 

In comparison, only little is known about \hd, and no thorough study of the star has been undertaken until now. The most interesting information comes from its environment: the star is surrounded by a bipolar nebula, NGC\,6164-6165, that displays a very symmetrical geometry.  Due to its anomalous chemical abundances, this nebula is thought to have formed through an eruption of the star, maybe similar to that of $\eta$ Carinae in the nineteenth century \citep{duf88}, making \hd\ a candidate Luminous Blue Variable object. 

The aim of our study is to assess the properties of \hd, in the visible and X-ray domains, and to compare it to the other two galactic Of?p stars. This paper is organized as follows. Section 2 describes the observations while Sections 3, 4 and 5 present the results of the available photometric datasets, the dedicated optical spectroscopic campaign and the \xmm\ observations, respectively. Section 6 gives our conclusions.

\section{Observations and Data Reduction}

\subsection{Visible domain}

Low-resolution spectroscopy of \hd\ was gathered between 2003 and 2006 via SMARTS (Small and Moderate Aperture Research Telescope System) at the CTIO 1.5m telescope equipped with the RC spectrograph. For each observation, three exposures of one to two minutes were combined, and the final signal-to-noise ratio (S/N) in the continuum was generally about 100. Three different gratings and/or settings were used: one covering the range 5650--6800\,\AA\ with a resolution of 2.8\AA\, the second 4000--4900\,\AA\  with a resolution of 2.0\AA, and the last 4065--4700\,\AA\  with a resolution of 1.6\AA. All the reductions were performed using the IRAF and {\sc midas} software. The normalization was done by fitting polynomials through carefully chosen continuum windows. The observations were relatively evenly distributed between the months of February and October. During the 4 years of the observing campaign, at least one blue+red spectrogram pair was taken each month of this visibility window. In total, we collected 53 red spectra and 56 blue ones.

Additional high-resolution spectra covering the whole optical range (3750--9200 \AA) were collected using the Fiber-Fed Extended Range Optical Spectrograph (FEROS), an echelle instrument mounted at the MPG/ESO 2.2-m telescope at La Silla. The resolution of FEROS is 0.1 \AA\ at 4800 \AA\ (thus R=48000); the exposure time was 15 min, leading to a signal-to-noise ratio of minimum 300. The spectra were all taken during the month of May (2003 May 24 and 25, 2004 May 6 and 2005 May 19). These data were reduced by the observers, E. Gosset and H. Sana, using an improved version of the FEROS context \citep[][]{san03} working under the {\sc midas} environment. After correcting for the blaze using flat fields, the different orders were normalized individually using polynomials of order 4--6. 

Other high-resolution observations can be found in the archives. A first one was taken with FEROS on 2005 June 26 and was kindly made available to us by the PI, J.-C. Bouret (S/N=200). Another one, taken on 1995 April 15 with University College London Echelle Spectrograph (UCLES), was found in the AAT archive. It consists of three exposures of 3--5min with S/N of about 140 that provide a spectrum covering the wavelength range from 3800 to 7400 \AA\ with a resolving power of 36000. The reduction of this observation was done in a classical way using the IRAF and {\sc midas} software. Finally, the star was also observed on 2002 February 27 with the Ultraviolet and Visual Echelle Spectrograph (UVES), in the framework of the UVES POP programme. The resolution was about 0.05\AA, and the signal-to-noise ratio $\sim$250. The reduced and merged UVES spectrum was downloaded from the public ESO database\footnote{http://www.sc.eso.org/santiago/uvespop }. Only a few selected regions were extracted from the UVES spectrum. All these spectra were corrected for the blaze using flat fields and then normalized using low-order polynomials. 

\subsection{X-ray range}

On 2001 Feb. 25 (Rev. 0223, PI R. Smith), \hd\ was observed twice consecutively with \xmm, for a total exposure time of 30ks. The star is associated with sources J163352.4--480640 and J163352.3--480641 in the 1XMM catalog. We retrieved the two datasets from the \xmm\ public archives, in order to perform a thorough analysis. For these observations, the three European Photon Imaging Cameras (EPICs) were operated in the standard, full-frame mode, except for the second pn dataset which was taken in the extended full frame mode. A medium filter was used to reject optical light. We processed these archival data with the Science Analysis System (SAS) software, version~7.0. After the pipeline chains (tasks {\sc emproc, epproc} and {\sc rgsproc}), the data were filtered as recommended by the SAS team: for the EPIC MOS (Metal Oxide Semi-conductor) detectors, we kept single, double, triple and quadruple events (i.e.\ pattern between 0 and 12) that pass through the \#XMMEA\_EM filter; for the EPIC pn detector, only single and double events (i.e.\ pattern between 0 and 4) with flag$=$0 were considered. To check for contamination by low-energy protons, we have further examined the light curve at high energies (Pulse Invariant channel number$>$10000, E$\gtrsim$10\,keV, and with pattern$=$0). No single, individual flare was detected during the observations but the background level was rather high and quite variable during the whole exposure. Further analysis was performed using the FTOOLS tasks and the XSPEC software v 11.2.0. 

\section{Photometry}

\hd\ was included in the ``New Catalogue of Suspected Variable Stars'', under the entry NSV\,7808. In this catalogue, it presents a V magnitude between 6.71 and 6.81~mag. However, the variability status of this star is still under debate. \citet{bal92} found a possible dimming of the star by about 0.01~mag over a few weeks time. Following \citet{van89}, the star rather presents a constant luminosity in $V$, although with a dispersion of 0.005~mag in April 1983, and possible small color variations ($\sim$0.002~mag). Comparing with older data, the same authors also noted that \hd\ might have been bluer and brighter in the late eighties than in 1960. Finally, \citet{van01} considers \hd\ as a possible candidate S\,Dor variable, but with only few indications for its S\,Dor status.

The three galactic Of?p stars were observed by the {\it Hipparcos} satellite\footnote{The database of the All-Sky Automated Survey (ASAS) also contains some data on \hd, but they are not usable: as is often the case for relatively bright stars, even the photometry flagged as good appears somewhat erratic, probably because of saturation problems. }. We have downloaded the individual photometry measurements from the {\it Hipparcos} database, including the {\it Tycho} data in the B and V filters and the {\it Hipparcos} broad band filter $H_p$. After discarding the measurements with non-zero flags to get rid of possibly problematic data, we have analyzed the photometry of \hd\ (=HIP\,81100, see Fig.\ \ref{phothip}). The star displays a H$_p$ magnitude varying between 6.79 and 6.83~mag, but the data are relatively sparse. Nevertheless, we have determined the reduced $\chi^2$ when fitting the data by a constant luminosity, calculated the autocorrelation for each dataset, and performed a period search. No significant change or periodicity was detected: the variable/S\,Dor status of \hd\ still awaits confirmation.

\section{Visible spectroscopy}

The full visible spectrum of \hd\ is shown in Fig. \ref{specferos}. Compared to recent, quiescent spectra of HD\,108, that of \hd\ presents appreciably weaker absorption lines of \niii$\lambda\lambda$4510-4534 and \heii. The faint \siii, \oii, and \cii\ emission lines seen in HD\,108 do not exist here, \oii$\lambda$4705 being rather in absorption.  The two \siv$\lambda\lambda$4088,4116 lines are however clearly in emission, whereas they were in absorption in the spectra of HD\,108. We may also note the important strength of the \hei$\lambda$6678 emission line, compared to \ha, in the spectrum of \hd. In addition, no P Cygni profile is seen for the \hei$\lambda\lambda$4471,4713 lines nor for the \hg\ or \hb\ lines (all are pure absorption). However, their profiles are not symmetric, with the red wing being much steeper than the blue one, indicating a possible contamination by some emission, most probably coming from the wind.

The spectral type of \hd\ appears rather early. In the high-resolution spectra, a visual comparison of the \hei$\lambda$4471 with \heii$\lambda$4542 lines and of \hei$\lambda$4026 with \heii$\lambda$4200 favors a O6 spectral type \citep[see Fig. \ref{specferos} and ][]{wal00b}. From the measurements of the equivalent widths (EW) of the former pair (see below), the high-resolution spectra again favors O6, whereas the low-resolution SMARTS data rather give O5.5, but close to the limit between the O5.5 and O6 types (see Fig. \ref{hei}). The star has been previously classified as O6 \citep{mcc76,pen96,gar77}, O6.5 \citep{wal72} and O7 \citep{con77,hum75}. These variations, if real, may indicate a similar behaviour as seen in HD\,108 and HD\,191612. However, these older spectral type determinations were reported more or less at the same epoch, and the differences could rather be due to different interpretations of the same spectrum. 

\subsection{Radial velocities and equivalent widths}

We investigated our 4 years of low-dispersion spectra in order to search for large, monthly to yearly variations. The radial velocity (RV) of the main spectral lines was measured by fitting a gaussian to the top (resp. bottom) of the emission (resp. absorption) lines. The EW was determined by integrating the profile in a window of about 8\AA\ centered on the rest wavelength (such a large window was necessary because of the low resolution of the SMARTS data). The mean of the RVs and EWs and dispersion around this mean are presented in Table \ref{tab: rvew}. Note that the RVs of \hei$\lambda$4471 appear systematically bluer than those of \heii$\lambda$4542 because of the contamination of the red wing of the \hei\ line.

The RV and EW of the diffuse interstellar bands (DIBs) near \ha\ and \civ\,\lo\,5812 were also measured. To get rid of possible remaining calibration problems, the observed RVs of these strong and rather narrow features were used to correct the RVs of the main lines, assuming the rest wavelengths from \citet{her95} and that the average DIB RVs of the FEROS data were the actual ones ($-$13.6\kms\ for the DIB at 5780\AA, $-3.4$\kms\ for the DIB at 6614\AA). However, only a slight improvement is seen in the dispersion of the RVs after this correction. 

The dispersion of the RVs appears rather constant among the measured lines (about 10\kms). The mean RVs from the high-resolution data (FEROS, UVES, UCLES) are compatible with the SMARTS values within the error bars (see Table \ref{tab: rvew} - note there is no dispersion for UVES and UCLES data since there is only one exposure). It is however worth noting that \hei$\lambda$4471 and \ha, two lines affected by wind emission, present a higher dispersion of the RVs in the high-quality FEROS datasets. Our measurements are also in agreement with the RVs found in the literature \citep{con77,aug85}. In addition, \citet{con77} mentioned the compatibility of their results with those of \citet{abt72} and their average value of the velocity is also similar, within the errors, to those of \citet{wes61,bus60}. 

From spectroscopic data, we have thus established a dispersion of 10\,km\,s$^{-1}$ (=$\sigma_{\rm RV}$) on the radial velocities of \hd.  Furthermore, no clear signature of binarity, like e.g. the presence of double-line profile, was detected in our data. Therefore, we conclude that no indication of binarity could be found on both the long-term (a few years) and short-term (a few days) ranges probed by the observations. This result agrees with the conclusion reached by \citet{con77} and \citet{gar80} on the constancy of the RVs\footnote{Using the same data as these authors, \citet{hut76} had suggested the star to be a binary of period 9d, but this was never confirmed later. However, a problem was identified with some of the photographic plates \citep{con77}.}.

However, it should be noted that the implications of the lack of significant RV variations in terms of the multiplicity of the star depend strongly on what assumptions we make about the orbital parameters of a putative binary. Indeed, unfavourable values of the orbital inclination, the mass ratio and the orbital period or a combination of several of these parameters could lead to rather small radial velocity variations. In this context, it is worth noting that RV variations similar to those found for HD\,191612 \citep{naz07,how07} would be below our detection threshold. The simple approach of \citet{gar80} enable us to evaluate the probability that we have missed a spectroscopic binary as a result of a low orbital inclination and for a given period and mass ratio. Taking the rather conservative assumption that the semi-amplitude of the radial velocity curve should be smaller than $2 \times \sigma_{\rm RV}$, and assuming a circular orbit and a mass of 55\,M$_{\odot}$ for the Of?p star, we find that there is a less than 1.7\% probability that \hd\ is a binary with a period of 100\,days or less and a mass ratio of 1. If we allow a mass ratio of 5 (i.e.\ a companion of spectral type $\sim$B2, the probability increases to 6 and 24\% for periods shorter than 15 and 100\,days respectively. Of course, the probabilities of missing a binary increase drastically if the orbit is no longer assumed to be circular and this would be especially relevant for a long-period binary system.

Looking at the EWs, three lines, \heii\,\lo\,4686, \hb, and \ha,  present clearly deviant dispersions, suggesting intrinsic variability. Fig. \ref{hadib} compares the RVs and EWs of \ha\ and the neighbouring DIB, presented with similar scales: the larger dispersion of the \ha\ EWs is obvious. We might note that these lines are the most sensitive ones to the wind. 

\subsection{Line profiles}

To investigate further the variability found previously, we performed another test using the Temporal Variance Spectrum \citep[TVS, ][]{ful96}. No large, significant variation was found, except for the aforementioned variable lines and \hei\,\lo\,6678. Values close to the significance threshold are however also found for \niii\,\lo\lo\,4634,4641 (see Fig. \ref{tvs}). 

Fig. \ref{specvariab} displays the line profile variations of \ha\ in the SMARTS data. Note that the changes occur on relatively short timescales, as underlined by our high-resolution data. In fact, two FEROS spectra taken on two consecutive nights do show the small scale variations in the \heii\,\lo\,4686 and Hydrogen Balmer lines (see right side of the same figure for \hb). Similar changes of the same lines are also found when comparing with the other high-resolution data. The variability is confirmed by looking at published tracings of the spectrum of \hd: in a spectrum taken in 1991 and reported in \citet{not96}, the \hb\ and \hg\ lines seemed to consist of two absorption components of equal strength in 1991, whereas the blue one is much stronger in our 2002-2003 high-resolution spectra\footnote{We might also note that from objective prism data, \citet{mcc76} reported that the \hb\ line of \hd\ was `filled in' but without a tracing, we can't compare their result with ours in detail.}. 

A period search was made on the RV and EW measurements using the techniques of \citet[see remarks of \citealt{gos01}]{hmm} and \citet{lk}. No significant period was detected, except maybe for \ha, where a peak at about 7d seems to slightly stand out above the noise. We also performed a 2D Fourier test on the SMARTS spectra themselves, in regions centered on the \ha, \hb, and \heii\,\lo\,4686 lines. The periodogram was then averaged on the wavelength interval (see the result in Fig. \ref{fourier}). This time, without ambiguity, a period of 7.031$\pm$0.003 is clearly detected for \ha. When folded with this period, the RVs and EWs of \ha\ actually follow a regular pattern (Fig. \ref{haphase}). This frequency is also present in the periodograms of \hb\ and \heii\,\lo\,4686, though with a reduced intensity. A short-term monitoring, preferentially done with a high resolution spectrograph, should better constrain the properties of this periodic phenomenon.

On the contrary, other lines, among which \hei\ and \ciii\,\lo\,4650, appear remarkably constant for an Of?p star. In that regard, the 1995 UCLES data could be a carbon copy of the latest FEROS data. Even older observations, taken by J.-M. Vreux in June 1974 and B. Westerlund in July and August 1974, show similar features, without any striking differences. The only reference to a possible change of these lines can be found in \citet{wes61}: `Practically all spectra have \heii\,\lo\,4686 in emission, most of them also \niii\,\lo\,4641 and a few also \ciii\ (?) \lo\,4651' - without having access to the actual plates, it is difficult to judge what happened that year, but maybe the crude photographic plates might be to blame. Without further information, it is difficult to judge the significance of this report; all we can say is that such a behavior is not seen in intensive, higher quality, subsequent data. 

\subsection{Physical parameters}

The visible spectrum can also be used to derive the physical parameters of \hd. For example, the line width reflects the projected rotation velocity of the star. Avoiding the variable Balmer Hydrogen and \heii\ lines and the contaminated \hei\ lines, we applied the Fourier method \citep[see][and references therein]{sim07} on the metal lines of \civ\,\lo\,5812 and \oiii\,\lo\,5592. This results in $v\sin(i)$ of 58\kms\ for the former and 45\kms\ for the latter, and we therefore conclude that $v\sin(i)\gtrsim45$\kms. This value is much lower than those found in the literature (200\kms, \citealt{con77b}; 92\kms, \citealt{pen96b}; 76\kms, \citealt{HSH}), but this is not surprising since the other methods do not permit one to disentangle the rotational broadening from other broadening mechanisms (i.e.\ macroturbulence, \citealt{sim07}).

An average of our FEROS spectra was fitted using CMFGEN \citep[for a description of the models and of the method, see also \citealt{mar05b,bou05}]{hil98}, as shown in Fig. \ref{cmfgen}. Note that, if the central part of the Balmer lines is contaminated by emission, the wings remain sufficiently clean to be used for gravity estimates. The best fit gives $T_{eff}=41000\pm2000$K, $\log(g)=4.0\pm0.1$, $R/R_{\odot}=15.0\pm2.5$ and $\log(L/L_{\odot})=5.75\pm0.1$ (assuming He/H of 0.08 and a distance of 1.38~kpc, see below). Such values are comparable to those of O5-6 V/III stars \citep{mar05}. 

The abundance of nitrogen appears to be N/H=$3\times10^{-4}$ (with an uncertainty of 40\%), which for a solar reference value of 7$\times10^{-5}$ \citep{asp05} corresponds to an overabundance of a factor 4 compared to the Sun. This overabundance indicates that the star is already chemically evolved, showing products of the CNO cycle at its surface. This result is compatible with abundance estimates in the surrounding nebula, thought to consist at least partly of material ejected by the star \citep[0.7 dex enrichment in N, see][]{duf88}. 

In addition, we estimated the mass-loss rate from the P Cygni profiles of an archival IUE spectrum. A rather high clumping factor ($f$=0.01) was required to prevent \niv\,\lo\,1720 (and to a lesser extent \ov\,\lo\,1371) from being too strong. This value was also found by \citet{bou05} in their study of two O4 stars. With this clumping factor, \niv\,\lo\,1720 and \siv\,\lo\lo\,1393,1403 are reasonably reproduced for a mass loss rate lower than 1--2$\times10^{-7}$\msol yr$^{-1}$. However, \nv\,\lo\lo\,1238,1242 and \civ\,\lo\lo\,1548,1550 remain too strong as long as $\dot M$ is larger than $\sim10^{-8}$\msol yr$^{-1}$. Given the  evidence for  non spherical emission in the star (see below), we refrained from going into too much detail in this analysis.  One can simply conclude that a mass-loss rate of  10$^{-7}$\msol yr$^{-1}$ (corresponding to $\dot M_{uncl}=10^{-6}$\msol yr$^{-1}$) is a conservative upper limit on the mass loss rate of \hd. Note that the prefered value for the terminal velocity amounts to 2600\kms. 

The fit appears rather good (see Fig. \ref{cmfgen}), except for two caveats. First, we note that all the observed \hei\ and \heii\ lines appears to be stronger than in a model with He/H=0.1, but reducing this ratio to 0.06 does not completely solve the problem. A similar difficulty was uncovered when fitting the spectrum of HD\,191612 \citep{wal03} and the origin of this discrepancy remains unclear at the moment (dilution by an hidden companion? contamination by emission?). Increasing the slope of the velocity field (the classical $\beta$ parameter) does not help to weaken systematically all He lines. Second, it is impossible to reproduce the visible emission lines with the mass-loss rate and terminal velocity estimated from the UV P-Cygni profiles. The Balmer and \heii\ emissions are narrow and cannot be explained (only) by a spherical wind emission as derived in the UV. Therefore, the wind should consist of two components: a spherical one (the only one CMFGEN is able to reproduce) and a non-spherical one where the narrow Balmer emissions arise (e.g. a disk).

\subsection{Summary}

Contrary to HD\,108 and HD\,191612, \hd\ does not display spectacular line changes for H, \hei\, and \ciii\,\lo\,4650. However, it does show small-scale variations of the \heii\,\lo\,4686 and Hydrogen Balmer lines. Looking closely at the \ha\ line, a period of 7.031d is detected on our low-resolution observations, but requires confirmation as the temporal sampling was not optimized for such short timescales.

The two other Galactic Of?p stars actually also display some small-scale changes, in addition to the main, large variations. For HD\,108, it can be easily spotted when comparing observations taken during one observing run (see e.g.\ Fig. \ref{108var}, reproduced from \citealt{naz04}). For HD\,191612, \citet{how07} found `small-amplitude variability ... on timescales longer than a few days'. It is not yet known if these changes are stochastic or periodic, as the observations were not optimized to detect such short-term variations.

Although of small amplitude, the phenomenon observed for \hd\ could still be related to those observed in the other Of?p stars. On the one hand, the variations of \hd\ could represent a scaled-down version of those observed for HD\,108 and HD\,191612, since these changes are similar in character. As the H and \heii\,\lo\,4686 variations were clearly the largest for these objects, it is possible that the \hei\ and \ciii\ changes would then remain undetected for \hd. The lower amplitude of the variability for \hd\ might be linked to a relaxation of the system following the eruption that gave rise to the bipolar nebula NGC6164-5, or could be related to a lower angle between the rotation and magnetic axis (in the case of the magnetic oblique rotator model, as proposed for HD\,191612\footnote{In this case, the 7d period would reflect the rotation period of the star. The magnetic field should therefore be weaker than for HD\,191612 since, although the star is old enough to have undergone eruptions, the field has not slowed the rotation to very long timescales as found for HD\,191612.}). On the other hand, it is equally possible that these small-scale changes are similar to the short-term variations seen in HD\,108 and HD\,191612. It would then remain to explain their periodicity in \hd, and see if the small-scale changes are periodic for the other two objects. In this case, larger variations could be present in \hd, but only with a much longer timescale than for HD\,108 since we see no significant changes of \hei\ and \ciii\ between 1974, 1995 and 2002-5 data.

%Note that the low-resolution data do not enable us to detect small-scale 
%variations: RVs variations such as those of HD\,191612 \citep{naz07} would have %been missed.

\section{X-ray data}

The X-ray emission  of \hd\ was first discovered with the $Einstein$ satellite. The star appears as source  2E1630.1$-$4800 in the $Einstein$ 2E catalog (Harris et al.\ 1994). It has an IPC count rate of 0.100$\pm$0.004 cts~s$^{-1}$ \citep{har94} or  0.114$\pm$0.004 cts~s$^{-1}$ \citep{chl89}. More recently, it  was observed by the \ro\ satellite, notably during the All-Sky Survey, where it appears as source 1RXS J163352.2$-$480643  (also known as 1RXP J163352.9$-$480635), with a PSPC count rate of 0.23$\pm$0.03 cts~s$^{-1}$. \ro\ has also observed \hd\ during  three pointed observations (collected under the id \# RP900379). The first observation was made in September 1992 during 3.35~ks, the second was taken in March 1993 during 10.14~ks and the last one in August  1993 for 5.36~ks. After downloading these data from the archives, we estimated PSPC count rates for these observations to 0.197$\pm$0.008 cts~s$^{-1}$, 0.189$\pm$0.004 cts~s$^{-1}$, and 0.203$\pm$0.006 cts~s$^{-1}$, respectively.

More recently, the star was observed by \xmm, whose high sensitivity permits a deeper analysis of its high-energy emission. The EPIC-MOS spectra of \hd\ were extracted over a circular region with radius 50\arcsec centred on the star; the background was extracted in the surrounding annular area of outer radius 75\arcsec. For EPIC-pn, the radius of the source region was limited to 37\farcs5 in order to avoid a nearby gap; we then use as background  a nearby circle devoid of sources. The EPIC spectra, binned to get a minimum of 10 cts per bin (i.e. S/N$\ge$3 in each bin), are shown in Fig. \ref{epicspec} where pn data is presented in green, MOS1 in black and MOS2 in red.

The spectrum appears thermal, and the Fe\,{\sc xxv} line at 6.7keV is present. We have adopted a distance of 1.38~kpc \citep[based on the membership in the Ara OB1a association, see][]{hum78} and an interstellar column density $N_H$ of 4$\times10^{21}$~cm$^{-2}$ \citep{dip94}. In the spectral modelling, we do not allow the absorption to go below this threshold. In addition, since both datasets gave consistent results, within the errors, we finally fit all data simultaneously. The results are very similar to those of HD\,108 and HD\,191612: fits using only one thermal component were unacceptable ($\chi^2>2$) whereas the sum of two absorbed\footnote{Since these two thermal components could arise in different regions of the wind, we allow them to be absorbed by independent column densities ($N^{\rm H}$).}, optically thin equilibrium plasma models ($mekal$, \citealt{kaa92}) is much better. In the latter case, two solutions with similar residuals are found: one with temperatures k$T$ around 0.6 and 2\,keV and negligible absorbing columns; the other with lower temperatures (0.2 and 1\,keV) and larger columns ($N^{\rm H}\sim 0.5$ and $1\times10^{22}$~cm$^{-2}$). The second is slightly better and also favored by the results of a fit by a differential emission-measure model (DEM, $c6pmekl$, \citealt{lem,sin}). The low-temperature peak is generally explained by instrinsic wind-wind shocks, whereas the high-temperature feature could be related to colliding winds in a binary or to shocks in a magnetically confined wind \citep{zhe07}. If we assume the latter origin, it is important to note that the importance of this second peak is much lower than for well-known magnetic system such as $\theta$\,Ori$^1$C or $\tau$Sco \citep{zhe07}.

As \hd\ is brighter and closer than the two other galactic Of?p stars, the noise on its X-ray spectrum is reduced. This helps to spot a small deficiency of the aforementioned model at low energies. The fit can be improved by the addition of a third $mekal$ component. Two solutions are again found, each corresponding to the two solutions mentioned above plus a third cooler component. Although the `hot' solution (k$T$ of 0.2, 0.6 and 2keV) presents a better $\chi^2$ than the `cool' one (k$T$ of 0.1, 0.2 and 1keV), it appears worse at low energies.

The results of these fits are reported in Table \ref{tab: specfit}, where the unabsorbed fluxes $f_{\rm X}^{\rm unabs}$ (in the 0.4--10.keV range) are corrected only for the interstellar absorbing column. For each parameter, the lower and upper limits of the 90\% confidence interval (derived from the {\sc error} command under XSPEC) are noted as indices and exponents, respectively. The normalisation factors are defined as $\frac{10^{-14}}{4\pi D^2} \int n_e n_{\rm H} dV$, where $D$, $n_e$ and $n_{\rm H}$ are respectively the distance to the source, the electron and proton density of the emitting plasma.

Using typical colors and bolometric corrections from \citet{mar06} and the optical properties of \hd\ ($V$=6.728, $B-V$=0.343, \citealt{mai04}), the bolometric luminosities amounts to $L_{\rm BOL}\sim 2\times 10^{39}$~erg~s$^{-1}$, a value in agreement with the results of model atmosphere fits. The X-ray luminosities $L_{\rm X}^{\rm unabs}$, evaluated from the EPIC data in the 0.5--10.0\,keV range and corrected for interstellar absorption, is $\sim2\times 10^{33}$~erg~s$^{-1}$, resulting in $\log (L_{\rm X}^{\rm unabs}/L_{\rm BOL})\sim-6$. Again, as for HD\,108 and HD\,191612, the value of this ratio is much larger (about 8 times here) than that of the `canonical' relation \citep{san06}. In this context, it is interesting to note that for the 2$T$ fit, only 30\%\ of the unabsorbed flux comes from the high-temperature component (in comparison, it is 25\%\ for HD\,108 and 27--35\%\ for HD\,191612): this means that the presence of a high-temperature component is not the only reason for the overluminosity. 

Finally, we also analyze the X-ray lightcurves. First, one ought to note that the \ro\ data and $Einstein$ count rates of \hd\ agree well with the spectral properties derived from the \xmm\ observations (Fig. \ref{cntnorm}). Second, lightcurves were derived from the \xmm\ datasets for a large range of energy domains and time bins. They were subsequently analyzed by $\chi^2$ and $pov$ tests \citep{san04} but no significant variation was found. It therefore appears that the flux of \hd\ at X-ray energies is constant, within the error bars, on short-term ranges (over the duration of an observation) as well as on timescales of decades.

\subsection{High resolution (RGS)}

The high-resolution (RGS) X-ray spectrum of \hd\ is presented in Fig. \ref{spectrum}. Although the signal-to-noise ratio is limited and in no ways comparable to that obtained for RGS spectra of closer objects, it is nevertheless possible to get some information from it. First, a global fit was undertaken: the lines were fitted by triangular profiles, which are particularly suited to get an idea of their width and their decentering (if any), while the residual continuum was fitted by a bremsstrahlung model. Only the amplitude of the triangles was allowed to vary from line to line, the overall shape being uniform. The best fit results in a line center position of $-195\pm707$\kms, a red edge at $1163\pm512$\kms, and a blue edge at $-2327\pm657$\kms, resulting in a FWHM of $1745\pm416$\kms. The X-rays lines are thus quite large and do not present significant blue/redshift, but their profiles are slightly skewed. In this context, it might also be interesting to note that the Ne\,{\sc ix}/Ne\,{\sc x} ratio for this O6 star fits nicely in the sequence found by \citet{wal07}. 

To get more precise information, we decided to focus on the O\,{\sc viii} $\lambda$\,18.97 Ly$\alpha$ line, the strongest feature in the spectrum that is rather free from blends with other lines. The line is broadened and has a FWHM around 2500\,km\,s$^{-1}$. A binned version of the data suggests that the line might display some structure (which could actually also be present on the Ly$\beta$ line, see below). We have attempted to fit the O\,{\sc viii} Ly$\alpha$ line with a exospheric line profile model following the formalism of \citet{Kramer}. The main assumptions are that the X-ray emission originates from material distributed throughout a spherical wind, above a radius $r \geq R_0 > R_*$ and that the hot plasma follows the bulk motion of the cool wind. Doppler broadening due to macroscopic motion hence provides the main source of line broadening. The line emissivity is assumed to scale as $\epsilon \propto \rho^2\,r^{-q}$ where the $r^{-q}$ term accounts for a radial dependence of the filling factor of the X-ray plasma. The free parameters of this model are thus $R_0$, $q$ (that we take equal to zero here) and $\tau_{\lambda,*} = \frac{\kappa_{\lambda}\,\dot{M}}{4\,\pi\,v_{\infty}\,R_*}$, the characteristic optical depth at wavelength $\lambda$. For the terminal wind velocity, we have first adopted $v_{\infty} = 2285$\,km\,s$^{-1}$ as derived by \citet{HSH}. The best fit to the unbinned line profile is obtained for $R_0 \simeq 2.25$\,$R_*$ and $\tau_{\lambda,*} = 0$. However, larger values of $\tau_{\lambda,*}$ are also possible for larger $R_0$. The fits to the binned profiles roughly confirm this picture, but bring up another (actually deeper) minimum in the $\chi^2$ contours around $R_0 \simeq 1.65$\,$R_*$ and $\tau_{\lambda,*} = 0.4$. While the unbinned data hence favour a flat-topped profile, the binned spectra rather suggest a skewed profile. In any case, we note that the fits suggest a rather broad line.

We have applied the same procedure to the O\,{\sc viii} $\lambda$\,16.01 Ly$\beta$ line. The best fit is obtained for $R_0 \simeq 1.85$\,$R_*$ and $\tau_{\lambda,*} = 0$, in reasonable agreement with the results obtained for the Ly$\alpha$ line. Note that the strong Ne\,{\sc x} $\lambda$\,12.13 Ly$\alpha$ line is probably blended with another line on its red side\footnote{There are a number of lines from various iron ions (e.g.\ Fe\,{\sc xxiii} $\lambda$\,12.193, Fe\,{\sc xvii} $\lambda$\,12.264 and Fe\,{\sc xxi} $\lambda$\,12.286) that could possibly be responsible for this blend. Given the presence of other strong lines of this ion, Fe\,{\sc xvii} is probably the best candidate.} and can thus not be fitted easily with our model.

We have repeated the fits of the O\,{\sc viii} $\lambda$\,18.97 Ly$\alpha$ and 
$\lambda$\,16.01 Ly$\beta$ lines with the same exospheric model and the same parameters as before, except for the terminal velocity where we adopted $v_{\infty} = 2600$\,km\,s$^{-1}$ instead (see section 4.3 above). The shape of the $\chi^2$ contours is essentially the same as found before with $v_{\infty} = 2285$\,km\,s$^{-1}$, but, as can be expected from the increase of the wind velocity, the minimum is shifted to somewhat lower values of $R_0$. In fact, the best fit to the unbinned Ly$\alpha$ line profile is now obtained for $R_0 \simeq 2.05$\,$R_*$ and $\tau_{\lambda,*} = 0$. The fits to the binned profiles now yield the lowest $\chi^2$ around $R_0 \simeq 1.45$\,$R_*$ and $\tau_{\lambda,*} = 0.0$. For the Ly$\beta$ line, the best fit remains at $R_0 \simeq 1.85$\,$R_*$ and $\tau_{\lambda,*} = 0$.

Over recent years, the He-like triplets, consisting of a forbidden ($f$), an intercombination ($i$) and a resonance ({$r$}) lines, of various ions have been used as plasma diagnostics for a number of O-type stars \citep[see e.g.][]{Leutenegger,Oskinova}. In fact the ${\cal R} = f/i$ ratio has been shown to be a sensitive diagnostic of the dilution of the UV radiation field in the line emission region \citep[e.g.][]{Porquet}. In the case of \hd, the only He-like triplet with a reasonable level of exposure is the Ne\,{\sc ix} triplet at $\lambda\lambda$ 13.447 ($r$), 13.548 + 13.551 ($i$) and 13.697 ($f$). However, even for this complex the data do not allow us to perform a quantitative fit of the line strengths. Still, the data show that the ${\cal R} \sim 1$ (see Fig.\,\ref{Nefir}). Hence this ratio must be below its collision equilibrium low-density limit of ${\cal R}_0 = 3.1$ \citep{Leutenegger}. Note also that the $f$ component could be blended with  un-resolved Fe\,{\sc xix} lines at $\lambda\lambda$\,13.73 -- 13.74. These features could lead to an overestimate of the actual strength of the $f$ line. The reduced strength of the $f$ component of the Ne\,{\sc ix} triplet is quite typical for O-stars \citep[see e.g.][]{Oskinova}. 
 
\section{Conclusions}

With its strong \ciii\,\lo\,4650 lines, \hd\ is a true Of?p star. However, it is unclear if it constitutes a perfect `twin' of HD\,108 or HD\,191612. This paper reports a first thorough, multiwavelength variability study of \hd.

In the visible domain, its photometry does not change much, and its S-Dor status can not be confirmed. The visible spectrum does vary, but the changes are limited to the main lines formed in the stellar wind, i.e.\ the Balmer Hydrogen  lines and \heii\,\lo\,4686. This variability occurs with very small amplitudes and an analysis of the changes in the \ha\ line reveals a periodicity of 7.031$\pm$0.003d. However, the \hei\ and \ciii\ lines seem constant. In addition, no large ($>$10\kms) variations of the RVs could be identified and no other typical signature of binarity (e.g.\ blended spectrum) was detected. Finally, model atmosphere fits yield parameters typical of an O5-6V/III star; they also suggest the presence of a non-spherical component to the stellar wind.

In the X-rays, the spectrum appears thermal in nature, with a dominant cool component (0.2keV), broad unshifted X-ray lines, and an order-of-magnitude overluminosity compared to normal O-type stars. It looks nearly identical to those of HD\,108 and HD\,191612, though with a larger flux (because of the smaller distance) and thus a better signal-to-noise ratio. No significant short-term or long-term variation of the X-ray flux could be brought to light. 

Three stars in our Galaxy share a very peculiar spectral characteristic, the presence of strong \ciii\ lines. Is \hd\ completely similar to the other two? Using only the X-ray results, the answer would be clearly yes. From the visible spectroscopy, however, the answer is unclear. The variability of \hd\ could either be related to the small-scale changes or to the large variations seen in HD\,108 and HD\,191912. Considering the former, additional data are needed to find if these small-scale changes are truly periodic for the other Galactic objects, and therefore determine the origin of this short-term, small-scale variability. In this case, it also remains to be seen if \hd\ presents large variations of its H, \hei, and \ciii\ lines - from our dataset, this can only occur on very long timescales (likely even larger than the 55yrs of HD\,108). On the other hand, the small-scale changes of \hd\ could represent a scaled version of the phenomenon observed in HD\,108 and HD\,191912, thereby explaining the lack of variations of the \hei\ and \ciii\ lines. In this case, the 538d period of HD\,191612 and the 55yrs timescale of HD\,108 would here be replaced by a much shorter, $\sim$7d period. 

To ascertain the origin of this variability and constrain more its properties, additional data are clearly needed: short-term monitoring of HD\,108 and HD\,191612, long-term observations at high resolution and spectropolarimetry of \hd.

\acknowledgments

We wish to thank Jean-Marie Vreux and Bengt Westerlund for providing scans of 1970s photographic plates. YN and GR acknowledge support from the Fonds National de la Recherche Scientifique (Belgium), the PRODEX XMM and Integral contracts, and the visitor's program of the STScI. STScI participation in the Small and Moderate Aperture Research Telescope System (SMARTS) Consortium at CTIO is funded by the Director's Discretionary Research Fund. FM thanks John Hillier for constant help with his code CMFGEN. \\ 

{\it Facilities:} \facility{XMM-Newton (EPIC)}, \facility{XMM-Newton (RGS)}.

\clearpage

\begin{figure}
%\plotone{hipparcos_148937.ps}
\caption{Hipparcos photometry of \hd. \label{phothip}}
\end{figure}

\begin{figure}
%\plotone{specferos_148937.ps}
\caption{Visible spectrum of \hd\ (median combination of our four FEROS spectra). The most important lines are labeled and tickmarks are drawn at their rest wavelengths. \label{specferos}}
\end{figure}

\begin{figure}
%\plotone{rvew_heiheii.ps}
\caption{RVs and EWs of the \hei$\lambda$4471 and \heii$\lambda$4542 lines, and the ratio of the EWs. SMARTS data are presented by filled circles and for the right panel, FEROS by stars, UVES by a cross (shown at HJD+1000.), UCLES by an open triangle (shown at HJD+4000.). The limits for the spectral types come from \citet{con71,con77c}. \label{hei}}
\end{figure}

\begin{figure}
%\plotone{rvew_haanddib.ps}
\caption{RVs and EWs of \ha\ and the DIB at $\lambda$6613.62\AA, shown on the same scale (SMARTS data only). \label{hadib}}
\end{figure}

\begin{figure}
%\plottwo{tvsblue2.ps}{tvsblue3.ps}
\epsscale{0.5}
%\plotone{tvsred.ps}
\caption{TVS and mean low-resolution spectrum of \hd. \label{tvs}}
\end{figure}

\begin{figure}
%\plotone{variab_148937.ps}
\caption{Left: long-term variability of the \ha\ line in the 2006 low-resolution spectra (from month to month). Similar figures can be drawn for the other years of the campaign. Right: variability of the \hb\ line in our FEROS spectra (2003: dotted line and black solid line for observations taken on May 24 and 25, respectively; 2004: thin dashed line, 2005: thin solid line). \label{specvariab}}
\end{figure}

\begin{figure}
%\plotone{red_ha_four.ps}
\caption{Top: Periodogram derived from the SMARTS spectra covering the \ha\ line (see text). Bottom: Spectral window of the observations. \label{fourier}}
\end{figure}

\begin{figure}
%\plotone{ha_in_phase.ps}
\caption{RVs and EWs of \ha\ phased with a period of 7.031d ($T_0$ was arbitrarily chosen as HJD=2\,452\,680). \label{haphase}}
\end{figure}

\begin{figure}
\caption{Comparison of the average FEROS spectrum of some lines (solid black line) and their best fit CMFGEN model (dash-dot red line, for $\log(\dot M)=-$7.5 and the parameters listed in Sect. 4.3). This figure only appears in colors in the electronic version of the journal. \label{cmfgen}}
\end{figure}

\begin{figure}
%\plotone{hd108hb_var.ps}
\caption{The \hb\ line profile in the spectrum of HD\,108 appeared particularly variable during our observing run of 2001: the thin solid line shows the spectrum taken on Sept. 10, the thick solid line the spectrum taken on Sept 11, dotted lines the spectra taken on Sept. 13 and 15, and dashed lines the spectra taken on Sept. 18. \label{108var}}
\end{figure}

\begin{figure}
%\plotone{epic_148937.ps}
\caption{EPIC X-ray spectrum of \hd\ with the best-fit 3$T$ model. The EPIC-pn data (drawn in green) appear at higher ordinates that those of the two EPIC-MOS (shown in black and red for MOS1 and 2, respectively). This figure only appears in colors in the electronic version of the journal. \label{epicspec}}
\end{figure}

\begin{figure}
%\plotone{ctrateXnorm.ps}
\caption{Normalized count rate: the \ro\ and $Einstein$ count rates were divided by the count rate expected from the best-fit 2$T$ model. A value of 1 indicates that no flux change has occurred. \label{cntnorm}}
\end{figure}

\begin{figure}
%\plotone{rgshd148937.ps}
\caption{Binned (3 channels per bin), combined (RGS1+2, 1st and 2nd order) X-ray spectrum of \hd. The most important emission lines are labeled. Note that the spectrum was not smoothed. \label{spectrum}}
\end{figure}

\begin{figure}
%\plotone{NeIXfir.ps}
\caption{\label{Nefir} The Ne\,{\sc ix} $fir$ He-like triplet in the RGS spectrum of \hd. The rest wavelengths of the different components are indicated by the labels. The $f$ component is clearly suppressed compared to the collision equilibrium low-density value of ${\cal R}_0 = 3.1$.} 
\end{figure}

%\clearpage

\clearpage

\begin{sidewaystable}
\begin{center}
\caption{ Mean and dispersion of the measured RVs and EWs. \label{tab: rvew}}
\begin{tabular}{lclclll}
\tableline\tableline
Line & EW interval & RV$_{smarts}$  & EW$_{smarts}$  & RV$_{feros}$ & RV$_{ucles}$  & RV$_{uves}$ \\
& (\AA) & (\kms) & (\AA) & (\kms)& (\kms)& (\kms) \\
\tableline
\heii\,\lo\,4199.83 &4193.83--4205.83  & $-46.0\pm$12.2 & 0.495$\pm$0.054 & $-38.0\pm$1.4 & $-$40.3 & $-$41.3 \\
\hei\,\lo\,4471.512 &4465.512--4475.   & $-59.3\pm$10.4 & 0.325$\pm$0.034 & $-41.0\pm$8.1 & $-$30.7 & $-$51.3 \\
\heii\,\lo\,4541.59 &4535.--4548.18    & $-38.7\pm$7.5  & 0.683$\pm$0.035 & $-32.2\pm$2.0 & $-$32.0 & $-$33.1 \\
\niii\,\lo\,4634.25 &4631.--4644.25    & $-24.4\pm$7.3  &$-0.963\pm$0.067 & $-29.7\pm$4.6 & $-$31.3 & $-$26.7 \\
\niii\,\lo\,4641.02 &                  & $-37.4\pm$6.7  &                 & $-43.6\pm$0.3 & $-$50.2 & $-$44.3 \\
\ciii\,\lo\,4650    &4644.25--4653.40  &                &$-0.470\pm$0.037 &               &   &   \\
\heii\,\lo\,4685.682&4679.682--4691.682& $-8.8\pm$7.5   &$-0.761\pm$0.079 & $-18.3\pm$4.7 & $-$22.8 & $-$19.4 \\
\hb\,\lo\,4861.33   &4645.33--4877.33  & $-191.2\pm$11.9& 1.218$\pm$0.120 &$-182.0\pm$4.6 &$-$185.0 &$-$187.7 \\
\ciii\,\lo\,5695.92 &5689.92--5701.92  & $-34.7\pm$11.9 &$-0.205\pm$0.041 & $-22.1\pm$0.4 & $-$28.9 & $-$23.2 \\
DIB\,\lo\,5780.45   &5774.45-5786.45   & $-19.8\pm$11.0 & 0.515$\pm$0.045 & $-13.6\pm$0.5 & $-$15.1 & gap \\
\civ\,\lo\,5811.98  &5805.98--5817.98  & $-49.3\pm$13.6 & 0.196$\pm$0.035 & $-33.3\pm$3.6 & $-$29.6 & gap \\
\civ\,\lo\,5812 (cor.)&                & $-42.7\pm$8.6  &                 &               &  &  \\
\ha\,\lo\,6562.85   &6550.85--6574.85  & 3.0$\pm$12.6   &$-1.176\pm$0.237 & $-7.1\pm$7.6  & $-$2.3  & $-$12.4 \\
\ha (cor.)&                            & 2.2$\pm$12.9   &                 &               &  &  \\
DIB\,\lo\,6613.42   &6609.42--6617.42  & $-2.2\pm$10.4  & 0.106$\pm$0.020 & $-3.4\pm$1.1  & $-$8.5  & $-$5.2 \\
\tableline
\end{tabular}
\end{center}
\end{sidewaystable}

\clearpage
\begin{sidewaystable}
\begin{center}
\caption{ Best-fitting models and X-ray fluxes at Earth. 
\label{tab: specfit}}
\begin{tabular}{>{\footnotesize}l >{\footnotesize}c >{\footnotesize}c >{\footnotesize}c >{\footnotesize}c >{\footnotesize}c >{\footnotesize}c >{\footnotesize}c >{\footnotesize}c >{\footnotesize}c >{\footnotesize}l >{\footnotesize}c >{\footnotesize}c}
\tableline\tableline
Type & $N_1^{\rm H}$  & k$T_1$   & $norm_1$     & $N_2^{\rm H}$  & k$T_2$ & $norm_2$    & $N_3^{\rm H}$  & k$T_3$   & $norm_3$ & $\chi^2_{\rm \nu}$(dof) & $f_{\rm X}^{\rm abs}$ & $f_{\rm X}^{\rm unabs}$\\
& $10^{22}$~cm$^{-2}$ & keV   & $10^{-3}$cm$^{-5}$ & $10^{22}$~cm$^{-2}$ & keV   & $10^{-3}$cm$^{-5}$         & $10^{22}$~cm$^{-2}$ & keV  & $10^{-3}$cm$^{-5}$        &                               & \multicolumn{2}{c}{(in $10^{-12}$~erg\,cm$^{-2}$\,s$^{-1}$)}  \\
\tableline
\vspace*{-0.3cm}&&&&& \\
2T & $0.49_{0.46}^{0.51}$ & $0.23_{0.22}^{0.23}$ & $33.9_{28.7}^{39.4}$ &  $1.03_{0.957}^{1.12}$ & $1.34_{1.31}^{1.37}$ & $4.42_{4.24}^{4.61}$ & &&& 1.34 (1804) & 3.0 & 7.0\\
\vspace*{-0.3cm}&&&&&\\
3T cool & $0._{0.}^{0.06}$ & $0.08_{0.08}^{0.09}$ & $19.5_{16.7}^{36.5}$ & $0.55_{0.51}^{0.57}$ & $0.23_{0.23}^{0.24}$ & $38.4_{30.4}^{43.3}$ & $1.23_{1.13}^{1.33}$ & $1.37_{1.33}^{1.40}$ & $4.52_{4.32}^{4.73}$ & 1.23 (1801) & 3.1 & 8.6\\
\vspace*{-0.3cm}&&&&& \\
3T hot & $0._{0.}^{0.01}$ & $0.25_{0.24}^{0.26}$ & $1.57_{1.49}^{1.71}$ & $0.35_{0.31}^{0.39}$ & $0.63_{0.62}^{0.64}$ & $3.43_{3.10}^{3.74}$ & $0.53_{0.44}^{0.63}$ & $2.00_{1.90}^{2.11}$ & $2.25_{2.11}^{2.39}$ & 1.13 (1801) & 3.2 & 7.8\\
\vspace*{-0.3cm}&&&&& \\
\tableline
\end{tabular}
\end{center}
\end{sidewaystable}


\begin{thebibliography}{}
%\bibitem[ ()]{}
\bibitem[Abt \& Biggs\ (1972)]{abt72} Abt, H.A., \& Biggs, E.S. 1972, Bibliography of stellar radial velocities, Latham Process Corp.
\bibitem[Asplund (2005)]{asp05} Asplund, M. 2005, ARA\&A, 43, 481 
\bibitem[Augensen\ (1985)]{aug85} Augensen, H.J. 1985, MNRAS, 213, 399
\bibitem[Balona\ (1992)]{bal92} Balona, L.A. 1992, MNRAS, 254, 404
\bibitem[Bouret et al.(2005)]{bou05} Bouret, J.-C., Lanz, T., \& Hillier, D.~J.\ 2005, \aap, 438, 301 
\bibitem[Buscombe \& Morris\ (1960)]{bus60} Buscombe, W., \& Morris, P.M. 1960, MNRAS, 121, 263
\bibitem[Chlebowski et al.\ (1989)]{chl89} Chlebowski, T., Harnden, F.R.Jr., \& Sciortino, S. 1989, ApJ, 341, 427
\bibitem[Conti \& Alschuler(1971)]{con71} Conti, P.~S., \& Alschuler, W.~R.\ 1971, \apj, 170, 325 
\bibitem[Conti \& Frost(1977)]{con77c} Conti, P.~S., \& Frost, S.~A.\ 1977, \apj, 212, 728 
\bibitem[Conti \& Ebbets\ (1977)]{con77b} Conti, P.S., \& Ebbets, D. 1977, ApJ, 213, 438
\bibitem[Conti et al.\ (1977)]{con77} Conti, P.S., Garmany, C.D., \& Hutchings, J.B.  1977, ApJ, 215, 561
\bibitem[Diplas \& Savage\ (1994)]{dip94} Diplas, A., \& Savage, B.D. 1994, ApJS, 93, 211
\bibitem[Donati et al.\ (2006)]{donati} Donati, J.-F., Howarth, I.D., Bouret, J.-C., Petit, P., Catala, C., \& Landstreet, J.  2006, MNRAS, 365, L6
\bibitem[Dufour et al.\ (1988)]{duf88} Dufour, R.J., Parker, R.A.R., \& Henize, K.G. 1988, ApJ, 327, 859
\bibitem[Fullerton et al.\ (1996)]{ful96} Fullerton, A.W., Gies, D.R., \& Bolton, C.T. 1996, ApJS, 103, 475
\bibitem[Garmany et al.\ (1980)]{gar80} Garmany, C.D., Conti, P.S., \& Massey, P. 1980, ApJ, 242, 1063
\bibitem[Garrisson et al.\ (1977)]{gar77} Garrisson, R.F., Hiltner, W.A., \& Schild, R.E. 1977, ApJS, 35, 111
\bibitem[Gosset et al.\ (2001)]{gos01} Gosset, E., Royer, P., Rauw, G., Manfroid, J., \& Vreux, J.-M.\ 2001, MNRAS, 327, 435
\bibitem[Harris et al.\ (1994)]{har94} Harris, D.E., Forman, W., Gioa, I.M., et al. 1994, EINSTEIN Observatory catalog of IPC X-ray sources (2E catalog),  SAO HEAD CD-ROM Series I ($Einstein$), Nos 18-36
\bibitem[Heck et al.\ (1985)]{hmm} Heck, A., Manfroid, J., Mersch, G. 1985, A\&AS, 59, 63
\bibitem[Herbig(1995)]{her95} Herbig, G.~H.\ 1995, \araa, 33, 19 
\bibitem[Heydari-Malayeri \& Melnick\ (1992)]{hey92} Heydari-Malayeri, M., \& Melnick, J. 1992, A\&A, 258, L13
\bibitem[Hillier \& Miller(1998)]{hil98} Hillier, D.~J., \& Miller, D.~L.\ 1998, \apj, 496, 407 
\bibitem[Howarth et al.\ (1997)]{HSH} Howarth, I.D., Siebert, K.W., \& Hussain, G.A.J.\ 1997, MNRAS, 284, 265
\bibitem[Howarth et al.\ (2007)]{how07} Howarth, I.D., Walborn, N.R., Lennon, D.J., et al. 2007, MNRAS, 381, 433
\bibitem[Humphreys\ (1975)]{hum75} Humphreys, R.M. 1975, A\&AS, 19, 243
\bibitem[Humphreys(1978)]{hum78} Humphreys, R.~M.\ 1978, \apjs, 38, 309 
\bibitem[Hutchings\ (1976)]{hut76} Hutchings, J.B. 1976, ApJ, 203, 438
\bibitem[Kaastra\ (1992)]{kaa92} Kaastra, J.S. 1992, An X-Ray Spectral Code for Optically Thin Plasmas (Internal SRON-Leiden Rep., Version 2.0) 
\bibitem[Kramer et al.\ (2003)]{Kramer} Kramer, R.H., Cohen, D.H., \& Owocki, S.P.\ 2003, ApJ, 592, 532
\bibitem[Lafler \& Kinman\ (1965)]{lk} Lafler, J., \& Kinman, T.D. 1965, ApJS, 11, 216
\bibitem[Lemen et al.\ (1989)]{lem} Lemen, J.R., Mewe, R., Schrijver, C.J., \& Fludra, A. 1989, ApJ, 341, 474
\bibitem[Leutenegger et al.\ (2006)]{Leutenegger} Leutenegger, M.A., Paerels, F.B.S., Kahn, S.M., \& Cohen, D.H.\ 2006, ApJ, 650, 1096
\bibitem[Mac Connell \& Bidelman\ (1976)]{mcc76} Mac Connell, D.J., \& Bidelman, W.P. 1976, AJ, 81, 225
\bibitem[Ma\'{\i}z-Apell\'aniz et al.\ (2004)]{mai04} Ma\'{\i}z-Apell\'aniz, J., Walborn, .R., Galu\'e, H.A., \& Wei, L.H.  2004, ApJS, 151, 103
\bibitem[Martins et al.\ (2005)]{mar05} Martins, F., Schaerer, D., \& Hillier, D.J. 2005, A\&A, 436, 1049
\bibitem[Martins et al.(2005b)]{mar05b} Martins, F., Schaerer, D., Hillier, D.~J., Meynadier, F., Heydari-Malayeri, M., \& Walborn, N.~R.\ 2005b, \aap, 441, 735 
\bibitem[Martins \& Plez\ (2006)]{mar06} Martins, F., \& Plez, B. 2006, A\&A, 457, 637
\bibitem[Massey \& Duffy\ (2001)]{mas01} Massey, P., \& Duffy, A.S. 2001, ApJ, 550, 713
\bibitem[Naz\'e\ (2004)]{naz04} Naz\'e Y. 2004, PhD thesis, Universit\'e de Li\`ege
\bibitem[Naz\'e et al.\ (2001)]{naz01} Naz\'e, Y., Vreux, J.-M., Rauw, G. 2001, A\&A 372, 195
\bibitem[Naz\'e et al.\ (2004)]{naz04b} Naz\'e, Y.,  Rauw, G., Vreux, J.-M., \& De Becker, M.\ 2004, A\&A, 417, 667
\bibitem[Naz\'e et al.\ (2006)]{naz06} Naz\'e, Y., Barbieri, C., Segafredo, A., Rauw, G., \& De Becker, M.\ 2006, IBVS, 5693
\bibitem[Naz\'e et al.\ (2007)]{naz07} Naz\'e, Y.,  Rauw, G., Pollock, A.M.T., Walborn, N.R., \& Howarth, I.D.\ 2007, MNRAS, 375, 145
\bibitem[Nota et al.\ (1996)]{not96} Nota, A., Pasquali, A., Drissen, L., Leitherer, C., Robert, C., Moffat, A.F.J., \& Schmutz, W. 1996, ApJS, 102, 383
\bibitem[Oskinova et al.\ (2006)]{Oskinova} Oskinova, L.M., Feldmeier, A., \& Hamann, W.-R.\ 2006, MNRAS, 372, 313
\bibitem[Penny\ (1996)]{pen96b} Penny, L.R. 1996, ApJ, 463, 737
\bibitem[Penny et al.\ (1996)]{pen96} Penny, L.R., Gies, D.R., \& Bagnuolo, W.G.Jr. 1996, ApJ, 460, 906
\bibitem[Porquet et al.\ (2001)]{Porquet} Porquet, D., Mewe, R., Dubau, J., Raassen, A.J.J., \& Kaastra, J.S.\ 2001, A\&A, 376, 1113
\bibitem[Sana et al.\ (2003)]{san03} Sana, H., Hensberge, H., Rauw, G., \& Gosset, E. 2003, A\&A, 405, 1063
\bibitem[Sana et al.\ (2004)]{san04} Sana, H., Stevens, I.R., Gosset, E., Rauw, G., \& Vreux, J.-M. 2004, MNRAS, 350, 809
\bibitem[Sana et al.\ (2006)]{san06} Sana H.,  Rauw G., Naz\'e Y., Gosset E., \& Vreux J.-M. 2006, MNRAS, 372, 661
\bibitem[Sim\'on-D\'{\i}az \& Herrero\ (2007)]{sim07} Sim\'on-D\'{\i}az, S., \& Herrero, A. 2007, A\&A, 468, 1063
\bibitem[Singh et al.\ (1996)]{sin} Singh, K.P., White, N.E., \& Drake, S.A. 1996, ApJ, 456, 766
\bibitem[van Genderen\ (2001)]{van01} van Genderen, A.M.  2001, A\&A, 366, 508 
\bibitem[van Genderen et al.\ (1989)]{van89} van Genderen, A.M., Bovenschen, H., Engelsman, E.C., et al. 1989, A\&AS, 79, 263
\bibitem[Walborn\ (1972)]{wal72} Walborn, N.R. 1972, AJ, 77, 312
\bibitem[Walborn\ (2007)]{wal07} Walborn, N.R. 2007, {\it Multiwavelength Systematics of OB Spectra}, Joint Discussion 4, The Ultraviolet Universe, IAU XXVI General Assembly, Prague, ed. A.~I. G\'omez de Castro \& M. Barstow (Universidad Complutense de Madrid), in press (astro-ph/0610837)
\bibitem[Walborn \& Fitzpatrick(2000)]{wal00b} Walborn, N.~R., \& Fitzpatrick, E.~L.\ 2000, \pasp, 112, 50 
\bibitem[Walborn et al.\ (2000)]{wal00} Walborn, N.R., Lennon, D.J., Heap, S.R., Lindler, D.J., Smith, L.J., Evans, C.J., \& Parker, J.Wm. 2000, PASP, 112, 1243
\bibitem[Walborn et al.\ (2003)]{wal03} Walborn, N.R., Howarth, I.D., Herrero, A., \& Lennon, D.J. 2003, ApJ, 588, 1025
\bibitem[Walborn et al.\ (2004)]{wal04} Walborn, N.R., Howarth, I.D., Rauw, G., et al. 2004, ApJ, 617, L61 
\bibitem[Westerlund\ (1961)]{wes61} Westerlund, B. 1961, Arkiv for Astronomii, 2, 467
\bibitem[Zhekov \& Palla\ (2007)]{zhe07} Zhekov, S.A., \& Palla, F. 2007, MNRAS, in press (arXiv:0708.0085)
\end{thebibliography}
\end{document}